\documentclass[aip,apl,reprint,superscriptaddress]{revtex4-1}

\usepackage{mathptmx}
\usepackage{amsmath}
\usepackage{bm}
\usepackage{graphicx}

\usepackage{amsfonts}
\usepackage{amssymb}
\usepackage{amsthm}
\usepackage{mathtools} % for dcases
\usepackage{acronym}
\usepackage[normalem]{ulem}
\usepackage{siunitx}
\usepackage[dvipsnames]{xcolor}

\graphicspath{{./figures/}}
\newcommand{\figwidth}{3.375in}

\newcommand{\etal}{\emph{et al.}}

\newcommand{\half}{\frac{1}{2}}

\usepackage{color}

\begin{document}

\title{Manifestation of higher-order inter-granular exchange in magnetic recording media}

\author{Matthew O. A. Ellis}
\affiliation{Department of Physics, University of York, York, YO10 5DD, United Kingdom}
\affiliation{School of Physics and CRANN, Trinity College Dublin, Dublin 2, Ireland}

\author{Razvan.-V. Ababei}
\affiliation{Department of Physics, University of York, York, YO10 5DD, United Kingdom}

\author{Roger Wood}
\affiliation{ HGST, Western Digital Company, San Jose, CA, United States }

\author{Richard F. L. Evans}
\affiliation{Department of Physics, University of York, York, YO10 5DD, United Kingdom}

\author{Roy W. Chantrell}
\affiliation{Department of Physics, University of York, York, YO10 5DD, United Kingdom}

\date{\today}

\begin{abstract}
Exchange coupling between magnetic grains is essential for maintaining the stability of stored information in magnetic recording media. Using an atomistic spin model, we have investigated the coupling between neighbouring magnetic grains where magnetic impurity atoms have migrated into the non-magnetic grain boundary. We find that when the impurity density is low a biquadratic term in addition to the bilinear term is required to properly describe the inter-granular exchange coupling. The temperature dependence both terms is found to follow a power law behaviour with the biquadratic exchange constant decaying faster than the bilinear. For increasing grain boundary thickness the inter-granular exchange is lower and decays faster with temperature. Further simulations of a grain at a bit boundary show an unexpected energy minimum for in-plane magnetisation which can only be reproduced using a biquadratic exchange term.
\end{abstract}

\maketitle

The coupling of magnetic grains in recording media is an important challenge for future high-density magnetic recording devices. To achieve suitably high areal densities the grain size and pitch distance are being pushed to a few nanometers\cite{Weller2014,Araki2008}. 
The properties of magnetic recording media, including the dynamic behaviour during the recording process\cite{Brandt2012a} and the long-term thermal stability\cite{Greaves2009}, are dependent on the the inter-granular interactions comprised of long-range magneto-static and short-ranged exchange coupling. However, the origin and nature of this short-ranged exchange coupling until recently was unexplored. As the design of recording media becomes more complex an understanding of the nature and magnitude of the exchange coupling and the development of realistic representation of the exchange for use in recording models becomes increasingly important.

Initial experimental measurements of inter-granular exchange by Sokalski \etal\cite{Sokalski2009} simplified the system to two thin films of magnetic media separated by a non-magnetic inter-layer giving a well-characterized system to investigate the physics of the exchange coupling. From these and other experimental measurements\cite{Granz2012,Huang2014} the inter-granular exchange is observed to decay exponentially with grain boundary thickness and linearly with temperature.

A model for this interaction is proposed to be magnetic impurities present in the grain boundary which provides a channel for the grains to interact directly\cite{Evans2014b}. Simulations using the proposed impurity model by Evans \etal\cite{Evans2014b} on a multilayer system similar to Sokalski's showed the same exponential dependence on grain boundary thickness as observed experimentally and also extracted the temperature dependence of the inter-granular exchange parameter which is important for future heat assisted recording schemes.

In addition to its origin, the particular angular form of the inter-granular exchange is known to have a large impact on the modelled bulk properties of granular media and on the recording performance.\cite{Wood2015,Igarashi2015,Donahue1997} The inter-granular exchange is typically written in the form of a Heisenberg exchange, often referred to as 'bilinear'\cite{Atkinson2016,Peng2011}. As we show later the Heisenberg form does not necessarily provide the correct angular dependence of the energy or torque. We find that at high impurity density the energy is dominated by a  micro-magnetic exchange stiffness behaviour, where at the limit with a continuous magnetisation the exchange energy can be modelled using 
\begin{equation}
	E_\text{ex} = \frac{J_\text{eff}}{2} \phi_{ij}^2 ,
	\label{eq:quad_free}
\end{equation} 
where $\phi_{ij} = \text{acos}(\mathbf{m}_i \cdot \mathbf{m}_j)$ is the total change in magnetization angle across the grain boundary that separates the grains.
However, in the low impurity density regime more relevant to magnetic recording media  we find that our results, obtained using an atomistic model, are better described by including both a bilinear and biquadratic Heisenberg exchange term of the form

\begin{align}
E_{ij}  = & A_{ij} ( 1 - \mathbf{m}_i \cdot \mathbf{m}_j ) + B_{ij} \left[1 -(\mathbf{m}_i \cdot \mathbf{m}_j)^2\right]. \label{eqn:biquad}
\end{align}

where $A_{ij}$ and $B_{ij}$ are the bilinear and biquadratic exchange constants between grains $i$ and $j$ and $\mathbf{m}_i$ is the normalised grain magnetisation. 

Biquadratic exchange has previously been observed in magnetic multi-layers\cite{Kreines2012}, such as Fe/Cr/Fe,\cite{Ruhrig1991}. Analysis by Slonczewski\cite{Slonczewski1991} determined that the origin of the biquadratic term lies not in the fundamental quantum mechanics, which does permit higher-order exchange terms, but arises from the variation of the inter-layer exchange due surface roughness causing frustration. Slonczewski's model is based upon the two layers interacting via a bilinear exchange term and leads to a biquadratic exchange whose magnitude depends on the variation of the bilinear exchange due to plateaus at the interface. This is independent of the exact origin of the bilinear exchange, i.e RKKY and/or super-exchange. Further analysis by Demokritov \etal\cite{Demokritov1994} shows that the magnetic dipole interactions can also contribute to this effect.

\begin{figure}
\centering
\includegraphics[width=2.25in]{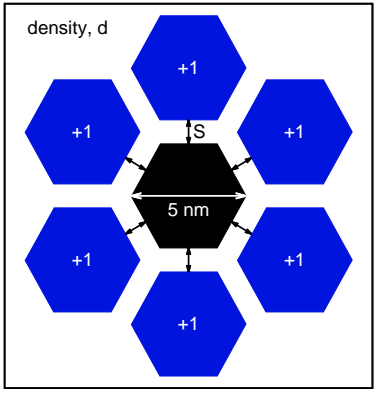}
\caption{Schematic showing the set-up used to calculate the inter-granular exchange. Each grain has a diameter of \SI{5}{nm} and the grain boundary thickness is given by the separation of the grains S. Initially the magnetisation of the outer grains are constrained to be $m_z=+1$.}
\label{fig:schematic}
\end{figure}

To explore the inter-granular exchange in recording media we employ an atomistic spin model\cite{Evans2014} based on localised magnetic moments. In contrast to reference \onlinecite{Evans2014b}, where a multi-layer is simulated, we model the recording media as a set of hexagonal grains; a central grain surrounded by 6 neighbouring grains of uniform size which is shown schematically in figure \ref{fig:schematic}. These grains are separated from each other by an non-magnetic grain boundary with a specified density of magnetic impurity atoms. These magnetic impurities provide a direct interaction channel without the need for long range interactions. On the atomic-scale the system is described by a Heisenberg exchange Hamiltonian
$\label{eq:ham}
  \mathcal{H} = - \half\sum_{i \ne j} J_{ij} \mathbf{S}_i \cdot \mathbf{S}_j,
$
where $J_{ij}$ is the exchange constant in terms of the unit vector spins $\mathbf{S}_i$. The summation for the exchange is carried out over nearest neighbours only.  Each atom has a magnetic moment of $\mu_s = \SI{1.44}{\mu_B}$ and the exchange constant is $J_{ij}=
\SI{5.6e-21}{J}$ which gives a $T_C \approx 1280$K. Each hexagonal 
grain has a corner-to-corner diameter of \SI{5}{nm} while the crystal structure is set as a
fcc lattice with unit cell size $a=\SI{3.54}{\angstrom}$. Periodic boundary conditions are employed in the $\mathbf{\hat{z}}$ direction to remove any surface effects and the
grain boundary thickness is varied by locating the grain centres further apart, this is noted by $S$ in Fig.~\ref{fig:schematic}.

Calculations are performed using the Constrained Monte Carlo (CMC) method~\cite{Asselin2010}, which fixes the direction of the magnetization but allows its magnitude to fluctuate. The CMC method generates joint moves of two individual spins such that the magnetization direction remains constant. A hybrid
constrained/unconstrained Monte-Carlo algorithm has previously been implemented
into the {\sc vampire} code\cite{Evans2014, Asselin2010} which allows us to fix the
magnetisation orientation for the grains whilst the spins within the
grain boundary are unconstrained and follow the usual Metropolis Monte-Carlo algorithm. 
Initially the surrounding grains are
constrained to be fixed in the $+z$ direction and the central grain is
constrained at different polar angles $\phi$
ranging from 0$^\circ$ to 180$^\circ$ in steps of 5$^\circ$, with $m_y = 0.0$. At each angle the macroscopic torque, $\boldsymbol{\tau}$, is computed for each grain from which the free energy can then be calculated\cite{Asselin2010}. The Monte-Carlo simulations are performed using
10,000 steps for equilibration and 10,000 steps to gather averages of the
properties for each separate constraint angle.

\begin{figure}
\centering
\includegraphics[width=\figwidth]{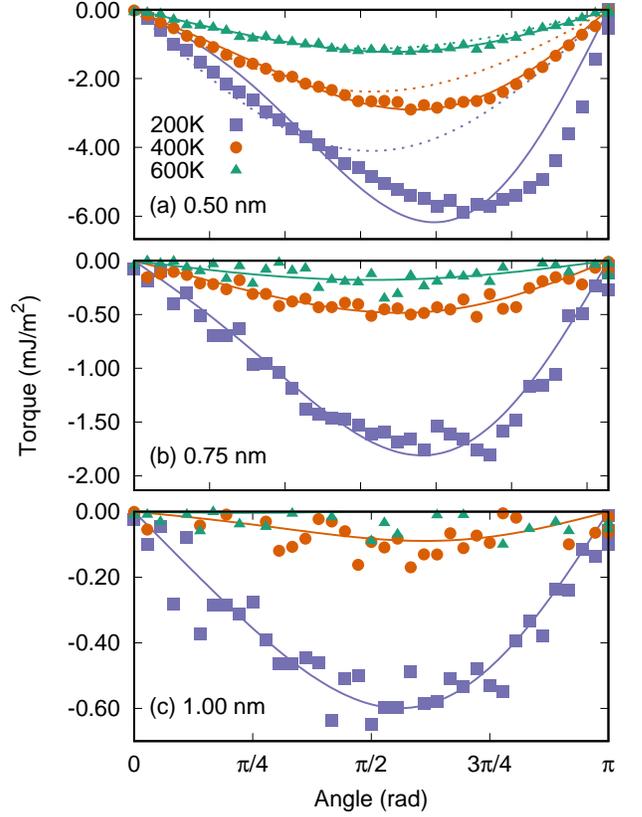}
\caption{ The torque acting on the central grain for separations (a) \SI{0.50}{nm}, (b) \SI{0.75}{nm} and (c) \SI{1.00}{nm} at 20\% density for \SI{100}{K}, \SI{300}{K} and \SI{500}{K}. Solid lines show a fit to the data using equation \ref{eqn:biquad} and the dotted lines in (a) show a fit using only the bilinear term.}
\label{fig:biquad_fit}
\end{figure}

To illustrate the non-Heisenberg angular variation of the inter-granular exchange energy, Fig.~\ref{fig:biquad_fit} shows the calculated torque acting on the central grain for three different temperatures and grain boundary thickness. Graphically we normalise the torque and energy by the contact area of the grains. In all cases there is a strong asymmetry in the torque which cannot be accounted for by using only a bilinear exchange term. An alternative is modelling the energy and torque using a micro-magnetic exchange stiffness approach which would yield a linear torque term: however this only agrees within a small angle limit. The solution we propose to this dilemma is to introduce a biquadratic term in the low density limit. Rather than fitting the free energy directly, which can be obtained from integrating the torque, we find that fitting the torque directly is more accurate since it is more sensitive to the angle. The inter-granular exchange is extracted by fitting

\begin{equation}
T(\phi) = -\frac{\partial E}{\partial \phi} = - A \sin\phi - 2 B \cos\phi \sin \phi. \label{eqn:torque}
\end{equation}

with $A,B$ as fitting parameters.
As shown in Fig.~\ref{fig:biquad_fit} the additional biquadratic exchange term provides an improved fit compared to only the bilinear term, which is shown as dotted lines in fig \ref{fig:biquad_fit}.(a). Importantly the biquadratic term explains the asymmetry in the results of the atomistic model predictions. The combination of bilinear and biquadratic terms describes the numerical results for the low exchange coupling regime but begins to break down with increasing inter-granular exchange corresponding to small grain boundary widths and low temperatures for example 200K in Fig.~\ref{fig:biquad_fit}(a).
This limits our biquadratic exchange model to low inter-granular exchange coupling  but this is in the regime that is of interest for experimental situations, i.e low impurity density and temperatures \SI{300}{K} and above.

% Exchange constants with temperature figure --------
\begin{figure}
\centering
\includegraphics[width=\figwidth]{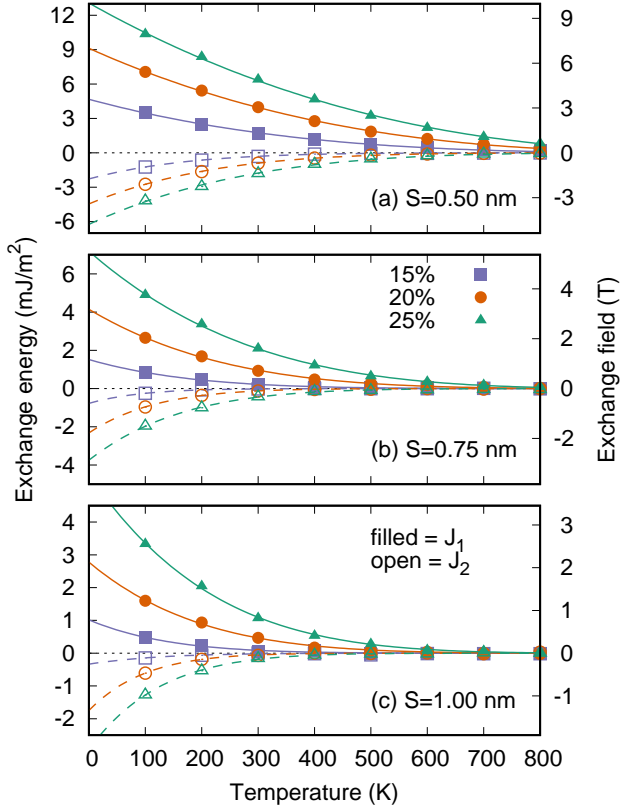}
\caption{The temperature dependence of the effective inter-granular exchange constants for (a) \SI{0.50}{nm}, (b) \SI{0.75}{nm} and (c) \SI{1.00}{nm} at low density. The bilinear exchange constant ($A_{ij}$) is shown as filled points and the biquadratic ($B_{ij}$) as open points and the lines are a fit to the data using equation \ref{eqn:power}.} \label{fig:J_v_temp}
\end{figure}
% End figure ---------------------------------------------

The variation of bilinear and biqudratic exchange constants, extracted from the fitting, with temperature is shown in Fig.~\ref{fig:J_v_temp}. The exchange energy can be expressed as an exchange field using $H_\text{exch} = A / M_s V$ using our reference saturation magnetisation of $M_s = \SI{6e5}{JT^{-1}m^{-3}}$ which is calculated from the magnetic moment per atom volume. There is a clear decay with increasing temperature which is best modelled using a power law expression. For both the bilinear and biquadratic exchange we fit
\begin{equation}
A(T) = A_0 \left(1 -\frac{T}{T_C}\right)^\alpha \label{eqn:power}
\end{equation}
to the data, where $A_0$ and $\alpha$ are fitting parameters and $T_C$ is held constant at the approximate Curie temperature (\SI{1200}{K}). 
This behaviour models the data well, as shown in Fig.~\ref{fig:J_v_temp} and we see that the biquadratic exchange decays faster than the bilinear term. This observation is consistent with previous simulations of the metamagnetic phase transition in FeRh based on a competition of bilinear and higher order (4-spin) terms\cite{Barker2015}.

Previous simulations have shown a linear temperature dependence of the exchange coupling\cite{Evans2014b} in agreement with earlier experimental measurements.\cite{Huang2014} In the present case the power scales with density and separation and so a situation may arise where the temperature dependence is close to linear. These results agree with reference \onlinecite{Evans2014b} in that at the write temperature for HAMR the inter-granular exchange will be negligible. However as the grain boundary thickness is reduced the exchange persists to increasingly high temperatures. We notice that the biquadratic term is always negative in our system which shifts the position of the peak torque to higher angles.

\begin{figure}
\centering
\includegraphics[width=\figwidth]{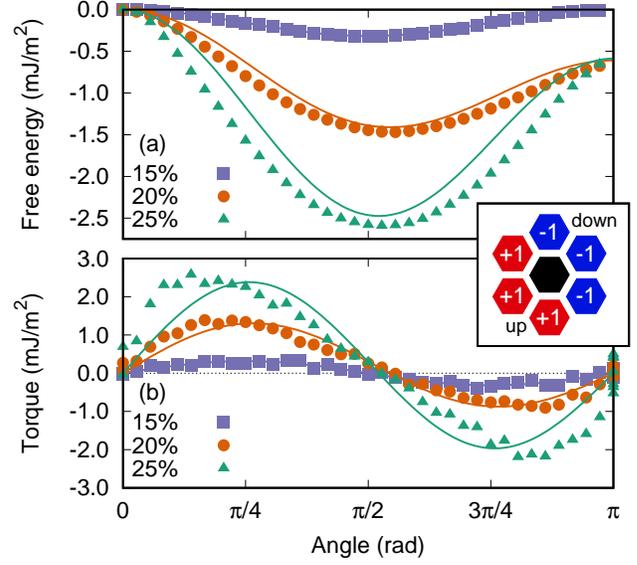}
\caption{Free energy (a) and torque (b) when the central grain has 3 neighbours constrained with $m_z = +1$ and 3 neighbours with $m_z = -1$ for 15\%, 20\% and 25\% density at \SI{300}{K}. The configuration is shown in the inset of (b). Lines show a fit using both bilinear and biquadratic terms as given in equation \ref{eqn:C2energy}.} \label{fig:C2fit}
\end{figure}

Finally we turn to the situation where the central grain lies at a bit boundary and the bits have opposite orientation. To represent this within our model we consider 3 neighbouring grains with their magnetisation up ($m_z=1$) and 3 with their magnetisation down ($m_z = -1$); a schematic of this configuration is shown as an inset on Fig.~\ref{fig:C2fit}.(b). In this configuration the central grain feels different torques forcing it to align in both orientations. Due to the symmetry of the bilinear term the energy contributions act against each other whilst the biquadratic terms are additive. In our model the exchange coupling arises from randomly sited impurity spins in the grain boundary, so there is a dispersion in the local exchange . This means that realistically the competing bilinear terms will not cancel completely but will depend on the difference of the bilinear constants. This leads us to a total energy of the form;
\begin{equation}
E = -\Delta A(1-\cos \phi) - B_\text{tot} \sin^2 \phi.
\label{eqn:C2energy}
\end{equation}
Where $\Delta A = A_\uparrow - A_\downarrow$ is the difference of the bilinear terms and $B_\text{tot} = B_\uparrow + B_\downarrow$ is to sum of the biquadratic terms while $\uparrow,\downarrow$ indicate the coupling to either neighbouring grain orientation.

Fig.~\ref{fig:C2fit} shows the free energy and torque calculated for the central grain for this configuration. The torque in panel (b) is fitted directly using equation \ref{eqn:torque} but now the coefficients represent the parameters in equation \ref{eqn:C2energy}, which  are then used to plot the energy in panel (a). There is a good agreement between the data and fit for the low densities and exchange values expected for recording media. At the higher densities the fit does not match the CMC data exactly but nonetheless provides a good qualitative representation of the key behaviour. The energy shows a slight mismatch as it is aligned with zero energy at $\phi = 0$ rather than the true minimum. In these examples there is a clear difference in the coupling to the different orientations; i.e $\Delta A \neq 0$ which leads to the asymmetry of the free energy. Without any anisotropy the central grain will minimise energy by having its magnetisation in-plane but if anisotropy is included there would be a preference to align with $-\mathbf{z}$ grains. If only the bilinear exchange is considered then it is clear the resulting $\sin\phi$ torque will not accurately represent the torque. Alternatively if a micro-magnetic exchange is considered where the torque can be approximated as $T= -A(\phi-\pi/2)$; this form fits well around $\pi/2$ but overestimates the torque close to 0 and $\pi$.

As discussed in the introduction, the origin of biquadratic exchange in multi-layers arises from interface roughness. The biquadratic exchange observed here may arise from a similar situation. The biquadratic exchange is dominant at low impurity density and so any clusters formed in the grain boundary will have varying length thus varying the energy of the domain wall that forms between each each grain. We note that a biquadratic term was not observed by Evans \etal~\cite{Evans2014b} for similar densities in a multi-layer structure where the torque is only observed to have a linear dependence with angle. However, the sample size was relatively small (an interface of 5nm$\times$5nm), so the biquadratic term was probably relatively small in this case.

To summarise, we have observed a manifestation of biquadratic inter-granular exchange in magnetic recording media. We have used an atomistic spin model to simulate a system with uniform hexagonal grains separated by an grain boundary in which magnetic impurities reside. Using a constrained Monte-Carlo method the torque and free energy are computed as a function of angle for the central grain. The angular dependence of the torque demonstrates the presence of a biquadratic exchange term. The temperature dependence of both the bilinear and biquadratic exchange follow a power law; with the biquadratic exchange decaying faster than the bilinear. Further simulations of a grain at a bit boundary again show the necessity of the biquadratic term. The origin of this manifestation is believed to be similar to Slonczewki's fluctuation mechanism for biquadratic exchange in magnetic multi-layer structures.

The authors gratefully acknowledge support from the Advanced Storage Technology Consortium. Work at Trinity College is supported by Science Foundation Ireland (Grant no. 14/IA/2624).

\bibliography{library.bib}

\end{document}